\documentstyle[12pt]{article}

\title{Non-Trivial Fixed Points of the Scalar Field Theory}
\author{
 K. Sailer,  
\\[2.812mm]
\hspace*{-8pt} Kossuth Lajos University, \\
H-4032 Debrecen, Hungary\\[0.2ex]
and 
\\[2.812mm]
W. Greiner
\\[2.812mm]
\hspace*{-8pt} Johann Wolfgang Goethe University, \\
D-60054 Frankfurt am Main, Germany
}

\begin{document}

\maketitle

\begin{center}
 KLTE-DTP/1996/2
\end{center}

\small
 
\abstract{
The phase structure of the  scalar field theory
 with arbitrary powers of the gradient operator   
  and a local non-analytic potential is investigated
by the help of the   RG in Euclidean space.
 The RG  equation for the generating function 
of the derivative part of the action is derived.
Infinitely many  non-trivial
fixed points of the RG transformations are found.
 The corresponding effective actions are unbounded from below and 
do probably  not exhibit any particle content. Therefore they do
 not provide physically sensible theories. 
}

\section{Introduction}
 
 As the
self-interacting scalar field with some continuous internal
symmetry is a basic ingredient of the
Standard Model and of GUT's, the phase structure of
 the one-component scalar field has also been of 
  a permanent
interest, in spite of the fact that it is only a toy model.
 It is well-known that scalar field theory exhibits 
 a Gaussian
fixed point \cite{Wi74}. For polynomial potentials that is
 a trivial IR fixed point corresponding to a free
scalar field. 
  The search for non-trivial fixed points
in  dimension $d=4$ gave negative
results (see \cite{HasNa88} and the literature cited there).
Perturbative renormalisability of scalar field
theory was proven by means of the RG 
using the scaling property of the operators at the Gaussian fixed
point \cite{Polch84,Wet90,Bal94}.
 A non-trivial fixed point was found in dimensions
 $2< d <4$, that however does not occur in dimension 4 \cite{HH86}.
A very interesting result has been obtained recently. Namely,
there are relevant directions at the Gaussian fixed point for
 analytic potentials \cite{HaHu95}. 
The existence of such relevant directions rises the question 
what is the low energy behaviour
of the theories defined at high-energy scale by one of the  relevant
potentials in the neighbourhood of the Gaussian fixed point.
 Therefore a renewed search for non-trivial fixed points has
 acquired actuality once again.

More recently it was argued in \cite{Mor96}
 that the non-trivial directions were found in
\cite{HaHu95} as a consequence of  the local potential approximation.
As far as one is looking for the RG flow in the coarse-graining
 direction,
any approximation neglecting irrelevant terms is justified and does not
spoil the search for the IR fixed point along a given RG trajectory.
Just the opposite is true if one tries to find UV fixed points with the
 help of the RG equations. Irrelevant terms negligible at a given scale
 become more and more important in the UV regime and reject the RG
 trajectory
from the UV fixed point. Therefore tracing back a RG trajectory
to the UV fixed point is practically impossible. How can one interpret
 the
situation that a fixed point has a relevant direction and the model
defined at some UV momentum scale lies  close to this fixed point
 in this direction?
Although it is not reasonable to ask for the UV fixed point, one can
 ask what is the IR effective theory for such a model.
That is what we are doing in the present paper without making use of
 the local potential approximation.

In the present paper we show that the one-component 
scalar field theory has infinitely many non-trivial fixed points
 in any dimension
$d>2$ and determine the fixed point actions and the scaling
operators in the neighbourhood of the non-trivial fixed points.
It is crucial for finding the non-trivial fixed points the enlargement
of the space of the scale dependent actions (Hamiltonians) considered
by (a) including terms with arbitrary powers of the gradient operator,  
and (b) local but non-analytic potentials and
 wave function renormalization.
Earlier investigations of the phase structure of the scalar field
theory used more restricted parameter space. The models 
considered did not contain terms with higher than the second power of
the gradient operator \cite{HH86,HaHu95}, \cite{Ara83}-\cite{Lan86}.
Below we show that the non-trivial
fixed point actions contain high powers of the gradient operator.
This may be the reason that they have been overseen before.

 The phase structure of the scalar field theory is investigated 
with the help of the Wegner-Houghton
equation \cite{WH73}. 
  As it is well-known, the  approach of Wegner and
Houghton enables one
to carry out renormalization  by  integrating out the
 high-frequency modes
$\Phi_q$ of the field $\Phi (x)$
 step by step in infinitesimally thin momentum shells,
$(k-\delta k , k)$. Making a quite general Ansatz with infinitely many
terms for the action $S_k$ at arbitrary scale $k$, the effect of the
 high
frequency modes can be incorporated into the change of its couplings
completely \cite{WH73,Var95,SP95,Po93}.  
 In this  procedure
the contribution of each momentum shell to the action can be calculated 
exactly
by making use of the small parameter $\delta k /k$.
With the help of the Wegner-Houghton equation  
 one can follow  the
evolution of the irrelevant coupling constants with the scale $k$, 
too. In the usual perturbative RG approach one only keeps track the
evolution of the very
limited number of 
 terms included 
 in  the bare action and cannot notice if some of the irrelevant
terms not included become relevant with decreasing scale $k$.

\section{RG Equations}

The  action at the momentum scale $k$  is assumed to have the form:
\begin{eqnarray}
\label{action}
  S_k &=&  \int d^d x  \left\{ 
 G_k (  \Phi  , -\partial^2 ) \Phi
   + U_k ( \Phi )  \right\} ,
\end{eqnarray}
where $G_k (  \Phi , -\partial^2 )$ is a local  functional
 of the field $\Phi (x)$ and an
analytic function of the derivatives:
\begin{eqnarray}
\label{deriv}
  G_k ( \Phi  , -\partial^2 ) = 
   \sum_{n=1}^\infty
   \sum_{r=0}^\infty  g_{nr} (k)   \Phi^{r+1} (x) 
         (-\partial)^{2n}  .
\end{eqnarray}
 Assuming that the system is enclosed in the 
finite volume $V_d$, 
we  rewrite   the
bare potential,
\begin{eqnarray}
\label{pot}
   {\cal U} &\equiv & \int d^d x U_k ( \Phi ) 
                = 
  \sum_{r=2}^\infty u_r (k) V_d^{-r}
    \sum_{q_1 , \ldots , q_r}^{\le k} \Phi_{q_1} \cdots \Phi_{q_r} 
    V_d \delta_{ q_1 + \ldots + q_r } 
\end{eqnarray}
and  the derivative part,
\begin{eqnarray}
   {\cal G} & \equiv & \int d^d x  G_k (  \Phi,
      -\partial^2 ) \Phi (x)
               \nonumber\\                  
       & =&  V_d^{-(r+2)}
           \sum_{n=1}^\infty \sum_{r=0}^\infty g_{nr} (k)
          \sum_{q_1 , \ldots , q_{r+2}}^{\le k}  
         \left(  q_1^2  \right)^n
         \Phi_{q_1} \cdots  \Phi_{q_{r+2} }
         \delta_{  q_1 + \ldots + q_{r+2} } .
\end{eqnarray}

The Wegner-Houghton equation can be written as \cite{WH73,HH86,SP95}:
\begin{eqnarray}
\label{WHeq}
   k \partial_k S_k &=&
             \frac{k}{2 \delta k} {\sum_p} '
                F_p K^{-1}_{p, -p} F_{-p}
      - \hbar \frac{k}{2 \delta k} {\sum_p} '
  \left( \ln  \frac{ K_{p,p'}  }{
    \left. K_{p , -p} \right|_{\Phi =\Phi_c} }
  \right)_{p, -p} ,
\end{eqnarray}
where $\Phi_c$ is either the vacuum expectation value $\Phi_0$, 
or any constant field configuration,
 the sum ${\sum_p }'$ is taken over the momentum shell
of the thickness $\delta k$  at $p^2 = k^2$, i.e. 
$ (1/V_d ) \sum_p ' \ldots =  \delta k \;  k^{d-1} \int
d\omega  (2\pi )^{-d} \ldots $ in the infinite volume limit,
 (with  the infinitesimal solid angle $d\omega$
in the $d$ dimensional momentum space)  and
$  F_p = \left(  \delta S_k / \delta \Phi_p 
          \right)_{\Phi_p =0}$,
$   K_{p, p'}  =  \left(
   \delta^2 S_k /  \delta \Phi_p \delta \Phi_{p'} 
     \right)_{\Phi_p =0} $,
where the subscript denotes that the Fourier amplitudes of the modes
in the momentum shell at $k$ have  to  be set to zero.
The first term on the r.h.s. of Eq. (\ref{WHeq}) is the tree level
 contribution occurring
 if $F_p \neq 0$.
 The second term on the r.h.s. of Eq. (\ref{WHeq}) is the
 one-loop contribution. The denominator in the argument of the
 logarithm
on the r.h.s. of Eq. (\ref{WHeq}) ensures that the effective action
$S_k$ takes vanishing value for $\Phi \equiv \Phi_c$
  for all values of $k$.
 Further on we shall work in the units $\hbar =1$.

The action considered has the general structure
 $ S_k ( \Phi_q )  =  V_d \sigma_k ( \phi_q )$
in terms of $\phi_q = \Phi_q /V_d $ and
 Eq. (\ref{WHeq}) takes the  form:
\begin{eqnarray}
\label{WH}
  k \partial_k \sigma_k ( \phi_q ) 
        &  = &
   \frac{k^d }{ 2 (2\pi )^d } \int d\omega 
    \left\lbrack
      V_d \; f_p k^{-1}_{p, -p}  f_{-p}
        - \left( \ln \frac{ k_{p, p'}  }{
                 \left.  k_{p, -p} \right|_{\phi = \phi_c }  }
          \right)_{p,-p}    \right\rbrack  
\end{eqnarray}
with $ f_p =\left.  \partial \sigma_k /\partial \phi_p
  \right|_{ \phi_p =0}$,
 and $k_{p,p'} = \left.  \partial^2 \sigma_k /\partial \phi_p
 \partial \phi_{p'}    \right|_{ \phi_p =0}$.

Our procedure of looking for the solutions of Eq. (\ref{WH}) is the 
following: we introduce 
 the generating functions for the derivative part and
 the potential, and derive partial differential
equations for them by an appropriate projection of the original
equation (\ref{WH})
and its second functional derivative:
\begin{eqnarray}
\label{WHdd}
\lefteqn{
   k \partial_k \frac{\partial^2 \sigma_k }{ \partial \phi_Q
              \partial \phi_{-Q}  }  
         }
       \nonumber\\    
  &=&
   \frac{k^d }{ 2 (2\pi )^d } \int d\omega 
     \frac{ \partial^2 }{ \partial \phi_Q \partial \phi_{-Q}  }   
      \left\lbrack
      V_d \; f_p k^{-1}_{p, -p}  f_{-p}
        -
     \left(  \ln \frac{ k_{p, p'}  }{
                 \left.  k_{p, -p} \right|_{\phi = \phi_c }  }
     \right)_{p,-p}   \right\rbrack  
\end{eqnarray} 
with $Q \neq  p$.
 The projector ${\cal P}$ we use
was introduced in \cite{HH86} by the definition
\begin{eqnarray}
 {\cal P} F &=& \left( \exp \left\{  x 
             \frac{\partial}{\partial \phi_0}  \right\}
                  F  \right)_{ \phi \equiv 0 } 
\end{eqnarray}
with the arbitrary functional $F$ of the field.
As shown in \cite{HH86} the projection of
any product of functionals equals to the product of the projections
of those functionals.

Let us define now the  generating function for the potential,
\begin{eqnarray}
\label{Vxk}
    V(x;k) &=&  \sum_{r=2}^\infty u_r (k) x^r ,
 \end{eqnarray}
and that for the derivative part of the action $S_k$,
\begin{eqnarray}
\label{GQxk}
    G (Q^2, x, k) &=& 2 \sum_{n=1}^\infty 
      Q^{2n}  \sum_{r=0}^\infty g_{nr} (k) x^r .
\end{eqnarray}
Here $V(x,k)$ is the generating function introduced in \cite{HH86}.

Let us
apply the projector ${\cal P}$ to both sides of Eqs. (\ref{WH}) and
(\ref{WHdd}).
Making use of the various steps of the projection described in the 
Appendix,
we find the following coupled set of partial differential equations
for the generating functions: 
\begin{eqnarray}
\label{WH1}
  k \partial_k V(x,k) = 
   - k^d \alpha \ln 
       \frac{   \partial_x G (k^2 , x,k)
               + \partial_x^2 V(x,k) }{  \left\lbrack
               \partial_x G (k^2 , x ,k)  
                       +  \partial_x^2 V(x,k)
             \right\rbrack_{x=x_c }  
            }   , 
\end{eqnarray}
\begin{eqnarray}
\label{WH2}
\lefteqn{
     k \partial_k 
   \partial_x G  (Q^2 , x, k)
                           }
          \nonumber\\
        &   =  & 
      - k^d \alpha \left\{
       \frac{ 
           \partial_x^3 G (k^2 , x,k)
                 }{ 
            \partial_x   G (k^2 , x ,k) 
   + \partial_x^2 V(x,k)
                              }
              \right.   
          \nonumber\\
     &   & 
        \left.  - \frac{
   \partial_x^2 G ( Q^2 , x, k) \left\lbrack \partial_x^2 G (k^2 ,x,k)
               + \partial^3_x V (x,k) \right\rbrack
      +  \left\lbrack \frac{1}{2}
 \partial_x^2 G (Q^2 ,x,k) \right\rbrack^2
             }{
       \left\lbrack  \partial_x G (k^2 ,x,k) + \partial_x^2 V (x,k)
       \right\rbrack^2 
             } 
        \right\} .
     \nonumber\\
\end{eqnarray}
with  $\alpha = \frac{1}{2} (2\pi)^{-d} \Omega_d$ ($\Omega_d$ the
entire solid angle in the $d$ dimensional momentum space).
Due to the particular choice of the action
 the tree level terms do not occur in these equations.  Therefore
Eqs. (\ref{WH1}) and (\ref{WH2})  are safe in the limit
 $V_d \to \infty$.

\section{Equations for  the Dimensionless Coupling Constants}

The equations for the dimensionless quantities can  be obtained by 
rescaling
the fields and the variable $x$ as
$   {\tilde \Phi}_q  = k^{ (d+2  )/2 } \Phi_q $,
and $   {\tilde x} = k^{ - (d-2 )/2 }  x$
that corresponds to ${\tilde x}_\mu = k  x_\mu$ and ${\tilde V}_d =
k^d V_d$ in coordinate space and leaves the dimensionless expression
$ \Phi_q / (V_d x) $ unchanged.
After this  rescaling we obtain
\begin{eqnarray}
  V ( x,k)  = k^d {\tilde V} ( {\tilde x} , k) ,
        & \qquad  &
 \partial_x^2 V( x, k)  = k^d k^{ -(d-2 ) }  \partial_{\tilde x}^2
      {\tilde V} ( {\tilde x} , k)  ,
\end{eqnarray}  
Requiring that ${\cal G}$ were dimensionless, we are lead to 
 $g_{10} = {\tilde g}_{10} $ and
\begin{eqnarray}
  G( Q^2 , x, k) &=& 2 k^{2 } \sum_{n=1}^\infty {\tilde Q}^{2n}
      \sum_{r=0}^\infty  {\tilde g}_{nr} {\tilde x}^{r+1} 
          =  k^{2  } {\tilde G} ( {\tilde Q}^2 , {\tilde x}, k ) .
\end{eqnarray}

The RG equations for the dimensionless generating functions take
then the following form:
\begin{eqnarray}
\label{WH1rs}
\left( k \partial_k - \frac{ d-2 }{2} {\tilde x} \partial_{\tilde x}
       +d \right) 
 {\tilde V} ( {\tilde x} , k) 
  &=&
 - \alpha \ln \frac{
               \partial_{\tilde x} {\tilde G} (1, {\tilde x}, k ) 
               + \partial_{ \tilde x}^2 {\tilde V} ( {\tilde x} , k)
                      }{
            \left\lbrack
               \partial_{\tilde x} {\tilde G} ( 1,  {\tilde x} , k)
                + \partial_{\tilde x}^2 {\tilde V} ( {\tilde x} , k)
             \right\rbrack_{ {\tilde x}_c } 
                       }  ,
      \nonumber\\
\end{eqnarray}
\begin{eqnarray}
\label{WH2rs}
\lefteqn{
    \left(  k \partial_k 
        - 2 {\tilde Q}^2 \partial_{ {\tilde Q}^2 } 
          - \frac{d-2 }{2} {\tilde x} \partial_{\tilde x} 
           +2
    \right) 
   \partial_{\tilde x} {\tilde G}  ( {\tilde Q}^2 ,  {\tilde x} , k ) 
         }       \nonumber\\         
    & =  &     
   - \alpha \left\{
     \frac{
  \partial_{\tilde x}^3 {\tilde G} ( {\tilde Q}^2 ,  {\tilde x} , k ) 
                                     }{
                 \partial_{\tilde x}  {\tilde G} ( 1,  {\tilde x}, k)
            +  \partial_{\tilde x}^2 {\tilde V} ( {\tilde x} , k)
                      }    \right.
                    \nonumber\\     
      &   &   \left.   - \frac{  
        \partial_{\tilde x}^2 {\tilde G} ( {\tilde Q}^2 ,{\tilde x},k)
           \left\lbrack 
          \partial_{\tilde x}^2 {\tilde G} (1, {\tilde x},k)
          + \partial_{\tilde x}^3 {\tilde V} ( {\tilde x},k)
           \right\rbrack
           +  \left\lbrack \frac{1}{2}
     \partial_{\tilde x}^2 {\tilde G} ( {\tilde Q}^2 , {\tilde x},k )
              \right\rbrack^2 
              }{
         \left\lbrack 
         \partial_{\tilde x} {\tilde G} ( 1, {\tilde x} , k) 
         + \partial_{\tilde x}^2 {\tilde V} ( {\tilde x} ,k) 
         \right\rbrack^2  
              }
            \right\} .
       \nonumber\\   
\end{eqnarray}

Now we generalize Eqs. (\ref{WH1rs}) and (\ref{WH2rs}) in two respects.
\begin{enumerate}
\item The above equations are valid, strictly speaking, only if the
      vacuum expectation value of the field is $ \phi_0 =0$.
       In the more general case $\phi_0 \neq 0$
      we have to take into account that the generating functions
      $V$, and $G$  depend on rather $x-x_0$ 
      than on $x$, and $x_0$ has the  dimensional scaling of $x$.
      Therefore the equations (\ref{WH1rs}) and (\ref{WH2rs}) must be
       modified 
     replacing ${\tilde x}$ by $z={\tilde x} - {\tilde x}_0$,
      and $x_c = x_0$ can be chosen.

      Due to the quadratic approximation of the action used by
      deriving the Wegner-Houghton equation \cite{WH73},
       the minimum $x_0 \neq 0$ of the action
      must be sufficiently close to zero, otherwise  the
      equation itself looses its validity.
\item Furthermore we assume the 
       validity of Eqs. (\ref{WH1rs}) and (\ref{WH2rs})
       also for generating functions $V$ and $G$ 
        non-analytic in the variable $z$.
       In the case of potentials having a singularity at their
       (absolute) minimum 
       at $z_c   =0$, we cannot keep the potential 
       at a fixed value in its minimum.
       Therefore we choose an arbitrary  point $z_c
       \neq 0$ for this
       purpose.
\end{enumerate}

All over this paper we shall only deal with dimensions $d>2$.

\section{Fixed Point Solutions}

The equations for the fixed points are obtained by setting zero the
derivatives of the generating functions with respect of the scale $k$
in Eqs. (\ref{WH1rs}), (\ref{WH2rs}).
We shall seek the fixed point solutions $V^* (z)$, 
$   G^* ( {\tilde Q}^2 , z )$ by  making the Ansatz
that the field dependent wave function renormalization can be
separated in the derivative part, i.e.
$ \partial_z G^* ( {\tilde Q}^2 , z ) =  H^* ( {\tilde Q}^2 ) h^*( z)$.

\subsection{Gaussian fixed point}

Assuming $V^* =const.$ and $h^* \equiv 1$, one easily finds the
fixed point solution
$H^* (  Q^2 ) =  H^*_0   Q^2 $, i.e. $G^* ( Q^2 , z) = H_0^* Q^2 z$
and the corresponding fixed point action
\begin{eqnarray}
S^* = - \frac{1}{2} H_0^* \int d^d x \phi (x) \partial_x^2 \phi (x) .
\end{eqnarray}
 The choice
$H_0^* =1$  can be made without loss of generality
 (rescaling of the field).

\subsection{Non-trivial fixed points}

We recognize that the fixed point equation obtained from
 Eq. (\ref{WH1rs})
has the solution  $ V^* =\frac{1}{2}  C_V \ln z^2 + V_0^*$,
 $h^* (z) = z^{-2}$ 
with $C_V = 2\alpha /d$ and \newline  $V_0^* = (\alpha /d^2)
\left( d-2  - d  \ln z_c^2 \right) $.
The logarithm on the r.h.s. of Eq. (\ref{WH1rs}) is only well-defined
for $b \equiv H^* (1)  - C_V \neq 0$. Then we obtain
a non-linear ordinary differential equation for the function
 $H^* (  {\tilde Q}^2 )$:
\begin{eqnarray}
\label{Hntr}
 {\tilde Q}^2  \frac{ d  H^* (  {\tilde Q}^2 ) 
               }{ d  {\tilde Q}^2 }
       &=&  \kappa   H^* (  {\tilde Q}^2 ) 
         - \nu \left\lbrack 
 H^* (  {\tilde Q}^2 ) \right\rbrack^2
\end{eqnarray}
with 
\begin{eqnarray}
    \kappa = 
    \frac{1}{2}  \left(  d + \frac{ 2 \alpha }{ b}  \right) ,
      \qquad
    \nu = \frac{ \alpha }{ 2b^2} .
\end{eqnarray}
It has the analytic solution:
\begin{eqnarray}
\label{Hstar}
  H^* (  {\tilde Q}^2 ) = \frac{\kappa}{\nu}  {\tilde Q}^{2 \kappa} 
         \left\lbrack C +  {\tilde Q}^{2 \kappa} \right\rbrack^{-1}
\end{eqnarray}
with 
\begin{eqnarray}
  \frac{\kappa }{\nu} = 
   2 C_V \frac{ 2\kappa}{d} \left\lbrack \frac{ 2\kappa}{d} -1 
                            \right\rbrack^{-2} ,
        \qquad
   C = \left( 3 -  \frac{ 2\kappa}{d} \right)
       \left(  \frac{ 2\kappa}{d} - 1 \right)^{-1} 
\end{eqnarray}
for positive integer  $\kappa \neq d/2$, $3d/2$.
The non-trivial fixed point solution is then given by
\begin{eqnarray}
  {\tilde  G}^* ( {\tilde Q}^2 ,  z ) 
      &= &
    - 2 C_V \left( 1 - \frac{d}{2\kappa} \right)^{-1}
      z^{-1}  {\tilde Q}^{2 \kappa} 
  \left\lbrack 3 - \frac{ 2\kappa}{d}  + 
       \left(  \frac{ 2\kappa}{d} -1 \right) 
      {\tilde Q}^{2 \kappa} \right\rbrack^{-1} ,
                \nonumber\\
  V^* (z) &=&
      \frac{1}{2} C_V \ln z^2 + V_0^* .
\end{eqnarray}

 The powers $\kappa = d/2$ and $3d/2$ must be excluded due
 to $H^* (1) \to \infty$. 
Analyticity and $H^* (0) =0$ are only satisfied for 
 $0 < \kappa $ integers.
Therefore there are infinitely many non-trivial 
 fixed points characterized by the
positive integers $\kappa \neq d/2$, $3d/2$.
 Eq. (\ref{Hstar}) is non-linear, therefore the
 linear combinations of the solutions corresponding to
 various values of $\kappa$ are not solutions.

It is worthwhile noticing that $\partial_z {\tilde G}^* = {\tilde Q}^2$
solves Eq. (\ref{WH2rs}) trivially. Then  Eq. (\ref{WH1rs})
leads to  the fixed point equation found in
\cite{HH86}. As discussed there and also in \cite{Mor94} (for $d=3$),
 a non-trivial IR fixed point
solution of Eq. (\ref{WH1rs}) exists
 for $2<d<4$  and,
 very probably, this fixed point
 does not occur for dimension $d=4$.
It is the constancy of the wave function renormalization 
the basic assumption for obtaining such a solution.

\section{Linearized RG Transformation}

As to the next we investigate the linearized RG transformations
in the neighbourhood of the fixed points in order to determine the
scaling operators.
 Let us write for the generating functions 
$ V (z,k) = V^* (z) + \delta V (z,k)$, and
$  G ( {\tilde Q}^2 , z,k ) = G^* ( {\tilde Q}^2, z ) +
\delta  G ( {\tilde Q}^2 , z,k )$ in the neighbourhood of a given fixed
point. In order to find the scaling operators at each of the fixed 
points
 we shall seek the eigensolutions of the operator $k \partial_k$
with the eigenvalues $-\lambda$ satisfying  Eqs. (\ref{WH1rs})
and (\ref{WH2rs}) in their linearized form. We make the separation
Ansatz for the solution:
\begin{eqnarray}
\label{eigen}
   \delta V (z,k)
     = \left( k/ \Lambda \right)^{-\lambda}
       \varphi ( z ) ,
                \qquad
  \partial_z  \delta G (  Q^2 , z, k) =
  \left( k/ \Lambda \right)^{-\lambda}
   \phi (  Q^2 ) \psi (z)  
\end{eqnarray}
with $\phi (0) =0$.
The positive, vanishing and negative eigenvalues $\lambda$ correspond
to relevant, marginal, and irrelevant directions, respectively,
at the given fixed point in the parameter space.

\subsection{Linearized RG equations at the Gaussian fixed point}

The linearized version of Eq. (\ref{WH1rs}) at the Gaussian fixed
point
is given by:
\begin{eqnarray}
\label{li1GFP}
\lefteqn{
   \left( k \partial_k - \frac{ d-2 }{ 2} z \partial_z + d
   \right)  \delta V (z,k)
        }
                    \nonumber\\
   & = & 
 - \alpha  
   \left\lbrack
   \delta \partial_z G ( 1,z,k)  - \delta \partial_z G (1, 0,k)  
   + \partial_z^2 \delta V (z,k)
   - \left. \partial_z^2 \delta V (z,k) \right|_{z=0}
   \right\rbrack ,
\end{eqnarray}
\begin{eqnarray}
\label{li2GFP}
  \left( k\partial_k
      + \alpha  \partial_z^2        
      - \frac{ d-2}{2} z \partial_z
        - 2 {\tilde Q}^2 \partial_{ {\tilde Q}^2 } 
        + 2
   \right)  \partial_z \delta G ( {\tilde Q}^2 , z, k) 
        &=&
          0  .
\end{eqnarray}   
For the eigensolutions (\ref{eigen}) we obtain
 the following equations from Eq. (\ref{li2GFP})
\begin{eqnarray}
\label{lipsiGFP}
    \left(  
       \alpha \partial_z^2  
    - \frac{ d-2}{2} z \partial_z  
      - \lambda + 2  -C 
    \right)  \psi (z)  =0 ,
\end{eqnarray}
\begin{eqnarray}
\label{liphiGFP}
   \left( 
  2 {\tilde Q}^2 \partial_{ {\tilde Q}^2 }
      - C  
    \right)  \phi (  {\tilde Q}^2 )  =0
\end{eqnarray} 
with the constant $C$. For $C =2n$ with $n=1,2, \ldots$ we obtain a
complete set of analytic solutions of Eq. (\ref{liphiGFP})
 satisfying $\phi (0) =0$:
\begin{eqnarray}
 \phi = \phi_0 ( {\tilde Q}^2 )^n
\end{eqnarray}
 with the constant $\phi_0$.
Let us write $ \psi (z) = \gamma Z (y)$, $z = \beta y$ and define the
constants $\beta$ and $\gamma$ as $\gamma = 4/(d-2 )$, $\beta^2
= \alpha \gamma $. Then we obtain the following ordinary
differential equation of second order for the function $Z (y)$ from
Eq. (\ref{lipsiGFP}):
\begin{eqnarray}
   \left\lbrack
    \partial_y^2  - 2 y \partial_y  
     + ( 2 - \lambda - 2n ) \gamma  
   \right\rbrack Z (y)  =0.
\end{eqnarray}
There is a complete set of analytic solutions characterized by
$ ( 2 - \lambda - 2n ) \gamma  = 2 n'$ with $n' =1,2, \ldots$,
namely the Hermite polynomials $Z (y) = H_{n'} (y)$. Thus we find
that the eigenvalues 
\begin{eqnarray}
   \lambda_{n n'} = 2 - 2n  - n' \frac{d-2}{2} 
\end{eqnarray}
are given by the integers $n$ and $n'$ and the corresponding
 eigensolutions are:
\begin{eqnarray}
\label{dGGFP}
 \partial_z \delta G
  &=& \left( \frac{k}{\Lambda } \right)^{ - \lambda_{nn'} }
      \frac{ 4}{ d-2 } \phi_0 (  {\tilde Q}^2 )^n H_{n'} 
            ( z/ \beta ) .
\end{eqnarray}

Inserting the solution (\ref{dGGFP}) in Eq. (\ref{li1GFP})
and introducing ${\bar \varphi} (z) = \varphi (z ) - C_2 =
 \gamma f (y)$ with $C_2 = 
 C_1 ( d- \lambda_{nn'} )^{-1}$, and
 $C_1 = \alpha  \left\lbrack \phi_0 \gamma H_{n'} (0)
+ \left. \partial_z^2 \varphi \right|_{z=0} \right\rbrack$,
 we find
\begin{eqnarray}
  \left\lbrack \partial_y^2 - 2y \partial_y 
         + ( d - \lambda_{nn' } ) \gamma 
   \right\rbrack f (y)
      &=& 
   - \alpha \phi_0 \gamma   H_{n'} (y)  .
\end{eqnarray}
By making use of the differential equation of the Hermite polynomials
we find the solution $f (y)= f_0 H_{n'} (y)$ with the constant $f_0$
determined via
\begin{eqnarray}
  d-2 + 2n &=& - \frac{ \alpha \phi_0 }{ f_0 } .
\end{eqnarray}  
The constant $C_2$ takes  the value $C_2 = - \gamma f_0 H_{n'} (0)$.
Then we can write the eigensolutions corresponding to the eigenvalue
$\lambda_{n n'}$ with $n$, $n'=1,2, \ldots$ as
\begin{eqnarray}
 \delta V &=&
 \left( \frac{k}{\Lambda } \right)^{ - \lambda_{nn'} }
 f_0 \frac{ 4}{ d-2  }
   \left\lbrack  H_{ n'} ( z/\beta )  - H_{n' } (0) 
   \right\rbrack  ,
\end{eqnarray}
\begin{eqnarray}
 \partial_z  \delta G &=&
  \left( \frac{k}{\Lambda } \right)^{ - \lambda_{nn'} }
   \frac{ - 4 f_0}{\alpha } 
   \left( 1 + \frac{ 2n}{d-2  } \right)
  (  {\tilde Q}^2 )^n H_{n'}  ( z/ \beta ) .
\end{eqnarray}
All these eigensolutions are irrelevant as $\lambda_{n n'} < 0$ for
the values taken by $n$ and $n'$.

It must be considered separately the case of
 the field independent
wave function renormalization $\psi (z) = const.$,
 corresponding formally to $n' =0$.
 There is no need then to introduce the constant $C$.
We can write $\psi (z) \equiv 1$ without loss of generality.
 The  equations (\ref{li1GFP}) and (\ref{li2GFP})
decouple and the eigensolutions for the kinetic part and for the
potential are completely independent.
We find the complete set of solutions 
$\partial_z  \delta G  = (k/\Lambda )^{-\lambda_n}
 \phi_0 ( {\tilde Q}^2 )^n$ belonging to the eigenvalues
$\lambda_n = 2  -2n$. The  solution $n=1$ is marginal, the higher
order derivative terms with $n >1$ are irrelevant. By introducing
${\bar \varphi} = \varphi - C_0 / (d-\lambda ) $
with $C_0 = \alpha \left. \partial_z^2 \varphi \right|_{z=0} $
 and writing
${\bar \varphi} (z) = \gamma f (y)$ with $z = \beta y$, we
obtain the following equation for the potential:
\begin{eqnarray}
  \left\lbrack
    \partial_y^2 - 2y \partial_y + ( d-\lambda ) \gamma 
  \right\rbrack f (y) =0 .
\end{eqnarray}
Let us now introduce the new variable $u=y^2$ and define ${\bar f} (u)
=f(y)$, then we find the confluent hypergeometric equation for the
function ${\bar f} (u)$:
\begin{eqnarray}
  \left\lbrack u \partial_u^2  + \left( \frac{1}{2} - u \right)
    \partial_u - ( a-1)
  \right\rbrack {\bar f} (u) =0
\end{eqnarray}
with $a-1 = \frac{1}{4} ( \lambda - d) \gamma$.

It has two independent solutions that can be chosen as
\begin{eqnarray}
\label{set1}
  {\bar f}_1 = G \left( a-1, \frac{1}{2} , u \right),
  \qquad 
  {\bar f}_2 = e^u G \left( \frac{3}{2} -a, \frac{1}{2}, -u \right) ; 
\end{eqnarray}
or
\begin{eqnarray}
\label{set2}
  {\bar f}_1 ' = F \left( a-1 , \frac{1}{2}, u \right) ,
    \qquad
  {\bar f}_2 ' = u^{1/2} F \left( a- \frac{1}{2} , \frac{3}{2}, u
\right)
\end{eqnarray}
with the confluent hypergeometric (Kummer) function $F(a,b,x)$ 
and the function $G(a,b,x) = \frac{ \Gamma ( 1-b) }{ \Gamma (a-b+1)}
F (a,b,x) + \frac{ \Gamma (b-1) }{ \Gamma (a) }
x^{1-b} F(a-b+1, 2-b, x)$. For $a = 1 - ( K /2 )$ with $K= 1,2,
\ldots$ the solutions ${\bar f}_1$ are the Hermite polynomials
as found in \cite{HH86}: $f (y) = 2^{-K} H_K (y)$ corresponding
to the eigenvalues $\lambda_K = d - (K/2) ( d-2 )$. So we recover
the well-known classification of the polynomial interactions
$\varphi (z) \sim \left\lbrack H_K ( z/\beta ) - H_K (0)
\right\rbrack$
 at the
Gaussian fixed point.  
 The quadratic potential $(K=2)$ is relevant.
The quartic potential $( K = 4 )$ is relevant, 
marginal, and irrelevant for $2<d<4$, $d=4$, and $d>4$, resp., higher
order terms in the potential are irrelevant for $d=4$. It was shown
in Ref. \cite{HH86} that the non-linear terms of the RG equation for
the potential  render the quartic potential also irrelevant for $d=4$.

 It has been observed more recently
\cite{HaHu95} that the non-polynomial
eigenpotentials 
\begin{eqnarray}
\varphi_a (z) = \gamma \left\lbrack
     F \left( a-1 , \frac{1}{2} , \frac{z^2 }{\beta^2} \right)  -1
                       \right\rbrack
\end{eqnarray} 
 corresponding to the solution ${\bar f}_1 '
= F ( a-1 , \frac{1}{2}, y^2 )$
 for $a_c \equiv - 2/(d-2) < a <0 $
i.e.  $\lambda_a = 2 + (d-2) a >0$ have a minimum. 
These potentials are relevant and behave
 asymptotically as $z^{2a -3} \exp \{ \beta^{-2} z^2 \}$.

 The various eigenvalues $\lambda$ correspond to
different directions in the parameter space around the Gaussian fixed
point. For $k \to 0$ and $\lambda <0$  the renormalization
 trajectory flows in the fixed point, representing a trivial IR  
fixed point for $d=4$. On the other hand, 
the trajectories for models with $a_c < a < 0$
move away from the fixed point. It represents  an UV fixed point
in this case, the corresponding models exhibit
 asymptotic freedom  \cite{HaHu95}, but it is not yet clear how these
models behave in the IR. As the equations for the potential and the 
derivative part of the action decouple completely, the couplings of
the  potential terms are independent of the couplings
of the derivative terms.

\subsection{Linearized RG transformation at the non-trivial fixed
              points}

As we have shown above, there are infinitely many fixed points
characterized by the positive integer  powers  $\kappa \neq d/2$  
 of the operator $\partial^2$. Restricting ourselves
to the solutions satisfying $\delta G ( 0, z, k) =0$, the linearized
 RG equations take 
the following form:
\begin{eqnarray}
\label{li1ntr}
\lefteqn{
\left( k \partial_k - \frac{ d-2}{2} z \partial_z
      +d  \right) \delta V  (z,k) 
            }  
              \nonumber\\
    & = & 
 - \frac{\alpha }{b} 
 \left\lbrack
  z^2 \partial_z \delta G (1,z,k) + z^2 \partial_z^2 \delta V (z,k)
  - z_c^2 \partial_z \delta G(1, z_c , k)
    - z_c^2 \partial_z^2 \delta V (z_c ,k)
  \right\rbrack ,
        \nonumber\\
\end{eqnarray}
\begin{eqnarray}
\label{li2ntr}
\lefteqn{
   \left( k \partial_k - 2 {\tilde Q}^2 \partial_{ {\tilde Q}^2 }
           - \frac{ d-2 }{ 2} z \partial_z  + 2 
   \right) \partial_z  \delta G ( {\tilde Q}^2 , z.k)
        } 
              \nonumber\\
    & = &
    - \frac{ \alpha }{b } 
      \left\{ 
    \left\lbrack 
      z^2 \partial_z^2  + 2 z \partial_z 
        + \frac{1}{b} H^* ( {\tilde Q}^2 ) z \partial_z 
    \right\rbrack \partial_z \delta G ( {\tilde Q}^2 , z,k)
           \right.
            \nonumber\\
       &   &  
      \left.     
 + \frac{2}{b} H^* ( {\tilde Q}^2 ) \left\lbrack
        1 + z \partial_z + \frac{1}{b }  H^* ( {\tilde Q}^2 )
           \right\rbrack
      \left\lbrack
      \partial_z  \delta G ( 1 , z.k)
           +  \partial_z^2  \delta V( z.k)
        \right\rbrack
       \right\}  .
\end{eqnarray}
Let us look once again for the eigensolutions corresponding to the
eigenvalue $-\lambda $ of the operator $k \partial_k$ in the form
given by Eq. (\ref{eigen}),
assuming $\phi (0) =0$ and making the Ansatz
\begin{eqnarray}
\label{ansatz}
   \phi (1) \psi (z) &=& - \partial_z^2 \varphi (z)  .
\end{eqnarray}

  Then we find the following  coupled set of
partial differential equations:
\begin{eqnarray}
\label{lipot1}
 \left( 
- \frac{ d-2}{2} z \partial_z - \lambda +d \right)
    \varphi (z)  &=& 0,
\end{eqnarray}
\begin{eqnarray}
\label{lider}
\lefteqn{
  \left( - 2 {\tilde Q}^2 \partial_{{\tilde Q}^2 }
      - \frac{ d-2}{ 2} z \partial_z 
      - \lambda + 2 
   \right)
     \phi (  {\tilde Q}^2 ) \psi (z)
        }
                 \nonumber\\
        &=& 
    - \frac{\alpha}{b} 
    \left\lbrack 
     z^2 \partial_z^2  + 2 z \partial_z  + 
     \frac{1}{b} H^* ( {\tilde Q}^2 ) z \partial_z 
    \right\rbrack
  \phi (  {\tilde Q}^2 ) \psi (z) .
\end{eqnarray}
Eq. (\ref{lipot1}) has  the solution:
\begin{eqnarray}
 \varphi (z) &=& \varphi_0 z^{2s} , \qquad
    s = \frac{ d-\lambda }{ d-2  } .
\end{eqnarray}
Then the wave function renormalization $\psi (z)$ is given by
\begin{eqnarray}
   \psi (z) &=& - \frac{ \varphi_0 }{ \phi (1) } 
   2s (2s-1) z^{2s-2} .
\end{eqnarray}
Finally we obtain from Eq. (\ref{lider}) the  equation
\begin{eqnarray}
    {\tilde Q}^2 \partial_{ {\tilde Q}^2 }
 \ln  \phi ( {\tilde Q}^2 )
          & = & 
      \rho  + \frac{\alpha}{b^2 } (s-1) H^* ( {\tilde Q}^2 ) 
\end{eqnarray}
with 
\begin{eqnarray}
   \rho &=& \frac{1}{2}
    \left\lbrack
    \frac{\alpha}{b} (2s-2) (2s-1) 
   -  \frac{ d-2}{2} (2s-2) 
     - \lambda + 2  \right\rbrack  .
\end{eqnarray}
Introducing the new variable $\xi = {\tilde Q}^{2\kappa}$, one can
 easily integrate this equation and find its solution:
\begin{eqnarray}
   \phi ( {\tilde Q}^2 ) \sim 
    {\tilde Q}^{2\rho}
    \left\lbrack 1 + C^{-1} {\tilde Q}^{2\kappa} \right\rbrack^{C_Q}
\end{eqnarray}
with $ C_Q = 2\kappa (2-\lambda )/(d-2)$.
The analytic solutions satisfying $\phi (0) =0$ are those with
\begin{eqnarray}
  \rho = C_Q \left\lbrack
     2 - \frac{d}{2} +
    \frac{d}{2\kappa} \left( C_Q \frac{\kappa -1}{\kappa} 
                              - 1 \right)
             \right\rbrack >0 ,
\end{eqnarray}
 $C_Q > 0$, and both $\rho$ and $C_Q$ integers.

One finds for
\begin{enumerate}
\item $d$ odd:
\begin{eqnarray}
  C_Q &=& 2 \kappa n,  \qquad n =1,2, \ldots 
        \nonumber\\
  \rho &=& \kappa n ( 4-d) + n d \left\lbrack 
       2n ( \kappa -1 ) -1 \right\rbrack ,
          \nonumber\\
  \lambda_n &=& 2- n (d-2) ;
\end{eqnarray}
\item $d$ even:
\begin{eqnarray}
   C_Q &=& \kappa n , \qquad n=1,2, \ldots
          \nonumber\\
  2 \rho &=& n \kappa ( 4-d) + n d \left\lbrack
     n ( \kappa - 1 ) - 1 \right\rbrack ,
           \nonumber\\
   \lambda_n &=&  2 0 n \left( (d/2) -1 \right) .
\end{eqnarray}
\end{enumerate}
Correspondingly
\begin{enumerate}
\item there are a relevant $(n=1)$, a marginal $(n=2)$,
and infinitely many irrelevant $(n>2)$ scaling operators for
dimensions $d=3$ and $d=4$;
\item there are a marginal $(n=1)$ and infinitely many irrelevant
 $(n>1)$ scaling operators for $d=6$;
\item for all the other dimensions $d$ all scaling operators are 
irrelevant.
\end{enumerate}

For the dimension $d=4$
there are no scaling operators analytic in ${\tilde Q}^2$ at
 the fixed point
with $\kappa =1$. The other fixed
 points with $\kappa >1$ exhibit a single relevant scaling operator
and can be considered as corresponding to critical theories. 
To each of those fixed points belongs a  critical
surface positioned perpendicularly to the relevant direction at the
given fixed point and the critical surfaces separate different phases.

Let us make a few important remarks.\\
$(i)$
The higher order derivative terms lead to non-localized interactions.
Causality is, however, not violated due to the analytic dependence
of the Lagrangian on the gradient operator \cite{Pai50}.
Unitarity depends on whether the real energy eigenvalue states are
all of positive norm \cite{Lee69}. It is, however, not a necessary
requirement for an effective theory to be unitary.\\
 $(ii)$
The fixed point actions
belonging  to all of the non-trivial fixed points are unbounded from
 below due to the logarithmic potential $V^* (z)$
 and rather probably none of them possesses any particle excitation.\\
$(iii)$
The sign of the derivative term of the fixed point action
 depends on $\kappa$. Let us consider the case of dimension $d=4$.
For $\kappa =1$ and $\kappa >6$ all Fourier modes give a positive
 contribution to the derivative part, whereas their contributions are
negative for $2< \kappa < 6$.
 The possibility of the existence
of such fixed points has already been argued in \cite{Gal83}.
 For such fixed point theories the 
vacuum corresponds to a periodic field configuration with the wave
 length
$2\pi/k$, but with vanishing amplitude due to the logarithmic 
potential.
\\
$(iv)$ It might happen that the highly non-linear RG equations have
 other
 fixed point solutions, not satisfying the separation Ansatz, but
 making more
 physical sense.

\section{Conclusions}

We have found infinitely many non-trivial fixed points for the
 one-component
 scalar field theory in the enlarged space of actions including
 derivative
 interactions and non-analytic potentials. The fixed point actions
found  are
however not really physical, they are
 unbounded from below and  do not support particle
excitations. If the RG trajectories starting at the Gaussian fixed
 point along 
a relevant direction flow towards them, the corresponding theories
 are not
sensical. In that case the only reasonable models are those being
 trivial.

\section*{Appendix: Derivation of the RG Equations }

The first and second derivatives of the action defined via
Eqs.  (\ref{action}), (\ref{deriv}), and (\ref{pot}) take the form:
\begin{eqnarray}
\label{der1}
  \frac{\partial \sigma_k }{ \partial \phi_Q }
     &=&
   \sum_{n=1}^\infty \sum_{r=0}^\infty  g_{nr} 
   \sum_{q_1 , \ldots , q_{r-1} }^{\le k } 
   \left\lbrack \left( Q^2 \right)^n + \left( q_1^2 \right)^n (r+1)
   \right\rbrack
    \phi_{q_1} \cdots \phi_{q_{r+1} } 
   \delta_{ q_1 + \ldots + q_{r+1} +Q}
           \nonumber\\
    & &
   + \sum_{r=2}^\infty u_r  
 \sum_{q_1 , \ldots , q_{r-1} }^{\le k } r
 \phi_{q_1} \cdots \phi_{q_{r+1} } 
   \delta_{ q_1 + \ldots + q_{r+1} +Q}  ,
\end{eqnarray}
\begin{eqnarray}
\label{der2}
  \frac{ \partial^2 \sigma_k }{ \partial \phi_Q \partial \phi_{Q'} }
      &=&
 \sum_{n=1}^\infty \sum_{r=0}^\infty  g_{nr}
 \sum_{q_1 , \ldots , q_{r} }^{\le k } 
     \left\lbrack  \left( Q^2 \right)^n +  \left( (Q' )^2 \right)^n
    + \left( q_1^2 \right)^n r
   \right\rbrack  (r+1)
    \phi_{q_1} \cdots
              \nonumber\\
      &  &  \cdots               
 \phi_{q_{r} } 
   \delta_{ q_1 + \ldots + q_{r}+Q +Q' }
           \nonumber\\
  & &
   + \sum_{r=2}^\infty u_r  
 \sum_{q_1 , \ldots , q_{r-2} }^{\le k } r (r-1)
 \phi_{q_1} \cdots \phi_{q_{r-2} } 
   \delta_{ q_1 + \ldots + q_{r-2} }  .
\end{eqnarray}
If we take the momenta $Q =p$, $Q' =p'$
 from the momentum shell at $|p|= |p'| =k$, 
we obtain $f_p$ and $k_{p, p'}$ resp. from Eqs. (\ref{der1}) and
 (\ref{der2})
by excluding the momenta of the shell from the sums, i.e. by changing
the upper limit of the sums over the momenta from $\le k$ to $< k$.
Furthermore we find
\begin{eqnarray}
\label{der1k}
\lefteqn{
 \frac{ \partial k_{p, p'} }{ \partial \phi_Q }
         }
    \nonumber\\   
 &=&
 \sum_{n=1}^\infty \sum_{r=0}^\infty g_{nr} 
    \sum_{q_1 , \ldots , q_{r-1} }^{< k } 
  \left\lbrack 
   2 k^{2n} + \left( Q^2 \right)^n +  \left( q_1^2 \right)^n (r-1)
 \right\rbrack r (r+1) 
    \phi_{q_1} \cdots
              \nonumber\\
    & &  \cdots       
 \phi_{q_{r-1} } 
   \delta_{ q_1 + \ldots + q_{r-1} +p + p' +Q   }
           \nonumber\\
 &   &
  + \sum_{r=2}^\infty u_r 
   \sum_{q_1 , \ldots , q_{r-3} }^{< k } 
    (r-2)(r-1)r
  \phi_{q_1} \cdots \phi_{q_{r-3} } 
   \delta_{ q_1 + \ldots + q_{r-3} +p+p' + Q   }  ,
\end{eqnarray}
\begin{eqnarray}
\label{der2k}
\lefteqn{
  \frac{ \partial^2 k_{p, p'} }{ \partial \phi_Q \partial \phi_{Q'} }
         }   
         \nonumber\\    
 &=&
 \sum_{n=1}^\infty \sum_{r=0}^\infty g_{nr} 
    \sum_{q_1 , \ldots , q_{r-2} }^{< k } 
  \left\lbrack 
   2 k^{2n} +  \left( Q^2 \right)^n +  \left( (Q')^2 \right)^n
 +  \left( q_1^2 \right)^n (r-2)
 \right\rbrack  \cdot
      \nonumber\\
   & & \cdot (r-1) r (r+1) 
    \phi_{q_1} \cdots 
           \phi_{q_{r-2} } 
   \delta_{ q_1 + \ldots + q_{r-2} +p+p' +Q +Q'   }
           \nonumber\\
 &   &
  + \sum_{r=2}^\infty u_r 
   \sum_{q_1 , \ldots , q_{r-4} }^{< k } 
   (r-3) (r-2)(r-1)r
  \phi_{q_1} \cdots \phi_{q_{r-4} } 
   \delta_{ q_1 + \ldots + q_{r-4} +p+p'+Q+Q'  }  .
         \nonumber\\
\end{eqnarray}

By means of the generating functions (\ref{Vxk}) and (\ref{GQxk})
we obtain:
\begin{eqnarray}
  {\cal P} \sigma_k &=& V(x;k) ,
      \\
  {\cal P} \frac{ \partial^2 \sigma_k}{ \partial \phi_Q 
  \partial \phi_{-Q}}
      &=&
  \partial_x G (Q^2 , x;k) + \partial_x^2 V(x ;k) ,
     \\
  {\cal P} k_{p, p'} &=&
      \left\lbrack
      \partial_x G (k^2 , x; k) + \partial_x^2 V (x; k)
     \right\rbrack \delta_{p+p'} \equiv {\cal A} \delta_{p+p'}  ,
       \\
   {\cal P} \frac{ \partial k_{p, p'} }{ \partial \phi_Q }
      &=&
    \left\lbrack
   \partial_x^2 G (k^2, x; k)  +\frac{1}{2} \partial_x^2 G (Q^2, x; k)
        + \partial_x^3 V(x; k)
      \right\rbrack \delta_{p+p'+Q}
          \nonumber\\
      &  \equiv &  {\cal B}  \delta_{p+p'+Q}  ,
         \\
  {\cal P} \frac{ \partial^2 k_{p, p'} }{ \partial \phi_Q
 \partial \phi_{Q'}}
  &=&  \left\lbrack
  \partial_x^3 G( k^2,x;k) + \frac{1}{2} \partial_x^3 G (Q^2, x;k) 
        \right.
        \nonumber\\  
  &  &
         \left.
 + \frac{1}{2} \partial_x^3 G ( (Q')^2, x;k)  + \partial_x^4 V(x;k) 
        \right\rbrack
       \delta_{p+p'+Q+Q'}
          \nonumber\\     
       &  \equiv & {\cal C}  \delta_{p+p'+Q+Q'} ,
            \\
    {\cal P}  (\ln k )_{p ,p'}  &=&
    \delta_{p+p'} \ln \left\lbrack 
         \partial_x G (k^2 , x; k) + \partial_x^2 V (x; k)
                    \right\rbrack
\end{eqnarray}

Applying the projector ${\cal P}$ on the r.h.s. of Eq. (\ref{WH}),
the tree level term vanishes due to ${\cal P} f_p =0$, as observed in
\cite{HH86}. Due to that the projection of  the second derivative
 of the tree level term
in Eq. (\ref{WHdd}) also takes the simpler form:
\begin{eqnarray}
  {\cal P} \frac{ \partial^2}{ \partial \phi_Q \partial \phi_{-Q} }
  \left(  f_p k^{-1}_{p.p'} f_{p'} \right)
    & = &
   {\cal P} \frac{\partial f_p}{ \partial \phi_{-Q} }
   {\cal P} k^{-1}_{p,p' } 
   {\cal P} \frac{\partial f_{p'} }{ \partial \phi_Q} 
           +
     {\cal P} \frac{\partial f_p}{ \partial \phi_{Q} }
   {\cal P} k^{-1}_{p,p' } 
   {\cal P} \frac{\partial f_{p'} }{ \partial \phi_{-Q} } .
       \nonumber\\
\end{eqnarray} 
However we find that ${\cal P} \left( \partial f_p / \partial \phi_Q
\right)  =0$ for any $Q \neq -p$, and therefore the tree level term
vanishes after acting with the projector ${\cal P}$ on the r.h.s.
of Eq. (\ref{WHdd}).

Making use of the matrix identity $\partial (kk^{-1} )_{p,p'}
 /\partial \phi_Q
=0$, we find
\begin{eqnarray}
   \frac{ \partial k^{-1}_{p,p'} }{ \partial \phi_Q }
       &=&
    - \left( k^{-1} \frac{ \partial k}{ \partial \phi_Q} k^{-1}
      \right)_{p,p'} 
\end{eqnarray}
and
\begin{eqnarray}
  \frac{ \partial^2 }{ \partial \phi_Q \partial \phi_{-Q}  }
   \left( \ln k \right)_{p,p'} 
  &=&
 \frac{1}{2} \left\lbrack
    k^{-1} \frac{ \partial^2 k}{ \partial \phi_Q \partial \phi_{-Q} }
  + \frac{ \partial^2 k}{ \partial \phi_Q \partial \phi_{-Q} }
    k^{-1}
         \right.
        \nonumber\\
     &   &  \left.
  - k^{-1} \frac{ \partial k}{ \partial \phi_{-Q} } k^{-1}
 \frac{ \partial k}{ \partial \phi_{Q} }
 -  \frac{ \partial k}{ \partial \phi_{Q} } k^{-1}
 \frac{ \partial k}{ \partial \phi_{-Q} } k^{-1} 
        \right\rbrack_{p,p'}
\end{eqnarray}
and then obtain
\begin{eqnarray}
   {\cal P} \frac{ \partial^2 }{ \partial \phi_Q \partial \phi_{-Q} }
          ( \ln k )_{p,p'} 
         &=&
    \left( \frac{ {\cal C} }{ {\cal A} } 
   - \frac{ {\cal B}^2 }{ {\cal A}^2 }  \right) \delta_{p+p'} .
\end{eqnarray}

Let us apply now the projector ${\cal P}$ on both sides of Eqs.
(\ref{WH}) and (\ref{WHdd}):
\begin{eqnarray}
\label{Vegy}
  k \partial_k V (x,k) &=& - \alpha k^d
 \ln \frac{\cal A}{ {\cal A}_{x_c} } ,
\end{eqnarray}
\begin{eqnarray}
\label{GVegy}
 k \partial_k \left\lbrack 
  \partial_x G ( Q^2 , x, k) + \partial_x^2 V(x,k) \right\rbrack
        &=&
   - \alpha k^d \left\lbrack
     \frac{ \cal C}{\cal A}  - \frac{ {\cal B}^2 }{ {\cal A}^2}
     \right\rbrack .
\end{eqnarray}

As far as $G(Q^2 , x,k)$ is assumed being analytic in the variable
 $Q^2$,
Eqs. (\ref{Vegy}) and (\ref{GVegy}) are consistent if and only if
the equation obtained from Eq. (\ref{GVegy}) by setting $Q^2 =0$,
\begin{eqnarray}
\label{cons}
     k \partial_k \partial_x^2 V &=& 
   - \alpha k^d \left\lbrack 
     \frac{ {\cal C}}{ {\cal A} } 
    -  \frac{ {\cal B}^2 }{ {\cal A}^2 } 
                \right\rbrack_{ Q^2 =0} , 
\end{eqnarray}
is the consequence of Eq. (\ref{Vegy}). Indeed, Eq. (\ref{cons})
can be obtained from Eq. (\ref{Vegy}) by taking the second partial
 derivative
of its both sides with respect of the variable $x$. 

Subtracting Eq. (\ref{cons}) from Eq. (\ref{GVegy}) we obtain
the coupled set of partial differential equations (\ref{WH1}) and
 (\ref{WH2})
for the generating functions.

\vskip 10pt
\noindent \large {\bf Acknowledgement}\normalsize
\vskip 10pt
\noindent

One of the authors (K.S.) is indebted to J. Alexandre, E. Branchina,
  J. Pol\'onyi, J. Rau,  and 
A. Sch\"afer  for the useful
 discussions.
 This work has been supported by the Hungarian Research Fund 
(OTKA, T 017311).

\end{document}